# Enhancing Cell Proliferation and Migration by MIR-Carbonyl Vibrational Coupling: Insights from Transcriptome Profiling

Xingkun Niu[a], Feng Gao[b], Shaojie Hou[c], Shihao Liu[b], Xinmin Zhao[a], Jun Guo[b,*], Liping Wang[b,*], Feng Zhang[a,b,*]

[a] Quantum Biophotonic Lab, Key Laboratory of Optical Technology and Instrument for Medicine, Ministry of Education, School of Optical-Electrical and Computer Engineering, University of Shanghai for Science and Technology, Shanghai 200093, China;

[b] Wenzhou Institute, University of Chinese Academy of Sciences, Wenzhou 325001, China;

[c] The School of Biomedical Engineering, Guangzhou Medical University, Panyu District, Guangzhou 511436, China

## Abstract

Cell proliferation and migration highly relate to normal tissue self-healing, therefore it is highly significant for artificial controlling. Recently, vibrational strong coupling between biomolecules and Mid-infrared (MIR) light photons has been successfully used to modify *in vitro* bioreactions, neuronal signaling and even animal behavior. However, the synergistic effects from molecules to cells remains unclear，and the regulation of MIR on cells needs to be explained from the molecular level. Herein, the proliferation rate and migration capacity of fibroblasts were increased by 156% and 162.5%, respectively, by vibratory coupling of 5.6 μm photons with carbonyl groups in biomolecules. Through transcriptome sequencing analysis, the regulatory mechanism of infrared light in 5.6 μm was explained from the level of signal pathway and cell components. 5.6-μm optical high power lasers can regulate cell function through vibrational strong coupling while minimizing photothermal damage. This work not only sheds light on the non-thermal effect on MIR light-based on wound healing, but also provides new evidence to future frequency medicine.

**Introduction**

Biomolecules can absorb the energy from their interacting electromagnetic (EM) waves or light due to the frequency-dependent resonant coupling [1]. Since the MIR band (4-25 μm) partially overlaps with the vibrational spectrum of biomolecules, therefore the MIR wave/light in principle could enhance the bond vibration via resonant excitation [2-3]. The photons absorbed through resonant coupling just likes a shot of pure energy to biomolecules, which not only alters their energy levels, or one can consider the excited molecules are dressed light/energy, more importantly also alters the reaction kinetics where they participate in, so-called nonthermal effect [4]. In this sense, MIR wave has been considered as the necessary energy band for life sustaining, which has been termed "the light of life".

In history, pioneers discovered the influence of EM radiation on biosystems. For instance, early in 1903, Niels Ryberg Finsen was awarded the Nobel Prize in physiology or medicine for his contribution to the treatment of skin diseases with light radiation.[5] Since then, "phototherapy" became a new nomenclature in medicine due to its simple, safe, and non-invasive characteristics. In 1967, Endre Mester, known as the father of laser therapy, first pioneered the field of laser biostimulation.[6-7] In the following years, the low-level laser therapy (also known as photobiomodulation, PMB) has been extensively studied and applied in medicine, e.g., oncology, neurological diseases, and medical cosmetology.[8-11] Einstein once said that "future medicine will be the medicine of frequencies." By coupling cytokines with different light frequencies, phototherapy can be applied to modify signal pathways and cells.[12-13] However, most of studies were in lack of in-depth insights into the molecular and physical mechanism, partially due to the utilization of wide range MIR light sources. For instance, using the full range of mid-infrared light (3-25 μm) can effectively promote the proliferation of human skin keratinocytes (HaCaT)[14] and proximal tubule cells (RPTCs)[15]. However, the working power needs to be reduced to a very low level to avoiding the thermal damages, and the radiation is slow and lasting.

From the absorption spectrum, water absorbs strongly around 3 μm and 6 μm, therefore the wide range MIR light including these two frequencies could induce strong thermal effect. For example, ceramic MIR emitters include the water resonant frequencies thus can cause a rapid increase in temperature of cells and the surrounding liquid environment. To avoid the thermal damage, MIR light power must be reduced (e.g., down to 0.13 mW/cm$^2$)[14] and the irradiation time was extended enough to exhibit the MIR potent effects on cells. In this sense, efficient phototherapy could not be achieved by wide range MIR light, but requires the precise MIR frequency.

In recent years, along with the gradual penetration of quantum theory to biomedicine and the rapid development of MIR light sources, a new nomenclature "vibrational strong coupling" (VSC) which requires a specific frequency, has been successfully transferred from physics, materials science to chemistry and biomedicine fields [16-19]. The current advances of VSC-based biomedicine can be classified into molecular, cellular, and animal levels, and investigations on both cellular and animal levels are based on the molecular level mechanism [4]. For examples, based on the VSC with -C=O groups, 5.6-μm laser has already been employed to significantly lower down the unwinding temperature of double-stranded DNAs (dsDNAs) [3] and improve the efficiency of polymerase chain reaction (PCR)[3], and to regulate the activity of $K^+$ ion channel thus influence neuronal signaling and zebra fish behavior[4].

Cell activities are fulfilled not by a single molecule or even a single reaction, but the synergetic reactions with many different biomolecules. For example, shining cells with a specific MIR frequency normally can excite many biomolecules with the same chemical group which can resonantly couple with, and these molecules could play different roles in many reactions, e.g., signal pathways. Meanwhile, biomolecular bond resonance also affects the structure and function of molecules, and also affects the binding and reaction properties of molecules, e.g., recognition of transcription factors or binding of growth factor receptors. For cells, transcriptome sequencing can disclose the information which genes are transcribed, and through enrichment analysis, we can find out which signal pathways or cell components play an important role in the influence of experimental conditions [20]. However, rare literature about this can be found so far [21].

Herein, we chose 5.6-μm laser to check out its quantum effect on fibroblasts for both proliferation and migration. The results showed that both the proliferation rate and the migration ability of fibroblasts were significantly promoted under a short-term irradiation (**Figure 1**). The specific effects of mid-infrared regulation on fibroblasts at the molecular level were analyzed by transcriptome sequencing. This work could pave the way to frequency medicine, e.g., precisely regulating cell functions through precise photon injection.

**Results and discussion**

The energy level of 5.6-μm light corresponds to the stretching vibration of carbonyl groups [4], which are abundant in cytokines, regulatory proteins, and is also an important component of DNA molecules. The vibrational coupling between 5.6-μm MIR light and carbonyl groups can give rise to coherent excitation of biomolecules[2-3]. In other words, coupling with the MIR light of a precise frequency equals to inject additional pure energy into biomolecules, which could enhance corresponding cell function, e.g., proliferation and migration. The quantum cascade laser can be used to emit a very narrow mid-infrared laser beam, The quantum energy of mid-infrared photons is low, and 5.6 μm is at the very low absorption of water, so the working power of our radiated laser in the experiment can be reached 1 W/cm$^2$, which is almost 1000 times powerful than the wide range MIR light sources[14]. However, the temperature increase was less than 0.1 °C (**Figure 2a**), suggesting a neglectable impact of thermal effects. The cell viability under different exposure times showed that there was almost no damage to cells compared to the control group during the 1-min MIR irradiation exposure (**Figure S1**).

To check out the thermal effect of 5.6-μm radiation, the temperature of cell media with different irradiation time was monitored. The results showed that the temperature increase was not detectable if the irradiation time was less than 4 min, and no more than 0.1 °C even if extending MIR exposure to 4 min. We also investigated the relative proliferation rate with different MIR irradiation doses. We used the ratio of the cell viability on day N after irradiation to the cell viability immediately after irradiation as the proliferation rate on day N, denoting the capacity to proliferate, and the ratio of the cell viability on day N after irradiation to the cell viability on day one before irradiation as the relative proliferation rate on day N, representing the change in the growth rate. In comparison, the relative proliferation rates at 120 h were 156%, 128.7% and 118.8% with irradiation time of 1 min, 30 s and 10 s, respectively. However, we found that the cell proliferation rate in the 2-min irradiation time group was less than 1, that is, cell proliferation was inhibited after 2-min irradiation time (**Figure 2b**). And 2min irradiation only increased the cell environment by 0.1 degrees Celsius, indicating that the proliferation inhibition effect was not entirely caused by thermal effects, including non-thermal effects caused by infrared light in 5.6 microns. Therefore, it is necessary to determine the appropriate working power and action time for MIR non-thermal effect regulation, that is, the total quantum energy emitted to the cell during the whole regulation time. Based on the above results, we chose 1-minute irradiation as the MIR regulation time for our design, because this configuration can also exclude thermal effects and cell viability is not impaired after irradiation. Clearly, the cells began showing stronger proliferative ability with 1-min MIR dose compared with the control group after 72 h. The proliferation rate increased by 268.5% at 120 h (**Figure 2c**), which can be clearly judged from the density statistics and the microscopic photos as well (**Figure 2d**). In combination, these results indicated a significant enhancement of cell proliferation.

We further investigated the irradiation effects on the cell migration ability by comparing the recovery of scratched areas, which was measured within 120 h after MIR irradiation. Results showed the difference of cell migration abilities between test and control groups can be distinguished since 24 h (**Figure 3a，Figure S2**). In comparison of the width of the scratch area at 120 h, the proportions for cells covering the scratch areas were 92.5% and 56.9% for the test group and control group, respectively (**Figure 3b**), indicating a significant MIR irradiation-induced enhancment on the migration ability of fibroblasts. Cell migration is a mechanical movement supported by the cytoskeleton, whose changes can be reflected by the tempospatial arrangement of actin.[22] With the fluorescence staining, the cytoskeleton of L929 cells can be quantitively analysized under microscopies. Starting from 24 h after 1-min 5.6-μm irradiation, cells became longer, exhibiting a relatively higher axial diameter ratio, compared with the control group (**Figure 3c-d**), indicating a remodeling initiation of the fibroblast skeleton. Overall, cell migration ability was significantly enhanced, which could be ascribed to the polarized rearrangement of actin filaments.

From the above experimental results, it can be concluded that the proliferation and migration of fibroblasts were enhanced significantly after MIR irradiation. To figure out the molecular mechanism, we conducted transcriptome sequencing on fibroblasts. By differentially transcribed genes after MIR irradiation, we analyze which biological processes they regulate cell proliferation and migration by altering, and analyze which cellular components or molecular functions they may affect. The differential generesults showed that 1230 genes were significantly upregulated and 2195 genes were significantly downregulated in fibroblasts after MIR irradiation (**Figure 4a**). From the bioinformatics aspect, gene ontology (GO) enrichment analysis of differentially expressed genes[23-24] revealed that the influence of differential genes related to MIR irradiation to enhance cell proliferation was localized in the nucleus, while the influence of differential genes related to enhance cell migration was localized in actin cytoskeleton, adhesion plaques, and collagen. (**Figure 4b**). In terms of biological processes, genes involved in cell cycle, DNA replication, and cholesterol synthesis are differentially transcribed (**Figure 4c**). In addition, the signal pathway enrichment analysis of differential genes showed that a large number of differential genes involved in TGF-β、ErbB and PI3k-Akt signaling pathways were transcribed after MIR irradiation, which can promote cell proliferation and migration by regulating the secretion of enzymes and the generation of growth factors.[25-27] Signaling pathways related to wound healing, muscle hypertrophy, and cholesterol synthesis further were activated as well, all suggesting the precise nonthermal regulation of 5.6-μm laser on the fibroblast proliferation and migration (**Figure 4d**).

We further investigated three typical upregulated genes involved in cell proliferation, i.e., Areg, which encodes the protein that activates intracellular signaling pathways such as MAPK and PI3K/AKT to stimulate fibroblast proliferation, migration and other functions;[28-30] Ccn1, which encodes a cell cycle protein that regulates cell proliferation; And previous experiments showed that PI3K/Akt and MAPK pathways were involved in the regulation of Ccn1 induced cell migration;[31-33] Serpine1, which is an important gene in epithelial cell migration and involved in multiple stages of skin repair, and its encoded protein PAI-1 can regulate cell migration by altering the proteolytic microenvironment around cells.[34-35]

With the real-time PCR (qPCR), we can see that the expression of these three genes was significantly increased at 24 h (**Figure 5a-c**). In combination with the experimental results of cell proliferation and migration, we can conclude that MIR irradiation significantly improved the expression of three genes. The corresponding expressed proteins encoded by these genes can activate the signaling pathways related to cell proliferation or migration, eventually enhanced the proliferation and migration of cells through regulating the fibroblast growth factors (**Figure 5d**). The promotion of cell proliferation and migration occurred only after MIR irradiation, and the expression of these three genes then gradually decreased to the normal level as the control group at 48 h and 72 h. This finding reflected that the activation effect of MIR irradiation on signaling pathways cannot last for a long time and will gradually recover. Since the effects of increased gene expression on cell migration require different time, the promotion of cell migration ability was obvious immediately after MIR irradiation, and became most significant at 72 h.

**Conclusions**

In summary, the proliferative and migratory abilities of epithelial cells were significantly improved by 1-min 5.6-μm irradiation. Transcriptome sequencing results revealed that the genes involved in the cell proliferation and migration were significantly upregulated, consistent with the qPCR results as well. The corresponding mechanism based on VSC between carbonyl vibrations and MIR light could be the state-of-the-art for understanding the effects of MIR irradiation on cell proliferation and migration. Such VSC-based light-molecule hybrids can generate higher polariton energy states, thereby modifies the chemical reactivity of biomolecules and their corresponding cell activities. To demonstrate the nonthermal effect of MIR-VSC on wound healing and repair, we must carefully choose the precise frequency of MIR irradiation to maximize the coupling strength, and meanwhile avoid the possible absorption by other molecules in biosystems, e.g., water. This work provides VSC-based frequency medicine as proposed by Einstein.

**Materials and Methods**

**Cell culture.** L929 mouse fibroblast cells were purchased from ATCC (Manassas, VA, USA). The cells were cultured in Dulbecco's Modified Eagle Medium (DMEM, Gibco, NYC, USA) supplemented with 10% fetal bovine serum (Pricella, Wuhan, China) and 1% penicillin/streptomycin (Beyotime, Jiangsu, China), and maintained at 37 °C in a 5% $CO_2$ environment with medium changed every 2-3 days.

**MIR laser emitter.** A pulsed quantum cascade laser (QCL, 176 mW average, MonoLux-56) was used to emit 5.6-μm MIR laser. The laser can be set in the emission wavelength range of 5.35-5.94 μm. A pulse-triggering frequency of 1-100 kHz, an output pulse width of 50-200 ns, and a rise/fall time of 5 ns. The output beam was nearly collimated with the lens inside the QCL package, with a divergence of 3 mrad in the vertical direction and 2 mrad in the horizontal direction. The laser was set to operate continuously at a wavelength of 5.6-μm, and the laser power was set with the distance of 100 mm from the laser aperture, whose power density was measured to be 1 $W/cm^2$.

**MIR Irradiation assay.** L929 cell suspensions in 96-well plates (Titan, Shanghai, China) were allowed to adhere overnight before laser Irradiation experiments. As proteins in the cell culture medium have strong absorption at 5.6 μm, the culture medium in the wells was replaced with PBS for the experiments with laser

Irradiation (including the control groups). The 96-well plate was placed under the laser aperture, and the wells were divided into experimental and control regions **(Figure S3)**. The cells in the experimental region were successively irradiated with 5.6-μm laser, with the unirradiated wells in the plate covered with a light shield during Irradiation. One well was irradiated at a time under ambient temperature. The cells in the control region of the 96-well plate received the same treatment as the experimental region, except for the laser Irradiation.

**Cell viability assay.** Cell viability was measured using the CCK-8 assay. Cells were seeded in a 96-well plate at a density of $5 \times 10^3$ cells/mL, with 200 μL of cell suspension per well. After laser Irradiation, cell viability was measured in both experimental and control groups every 24 h. 10 μL of CCK-8 solution (Yeasen, Shanghai, China) was added to each well, and the plate was incubated for 3 h before measuring the absorbance at 450 nm using a microplate reader (Synergy H450, BioTek, Vermont, USA).

**Cell migration assay.** Cells were seeded in a 96-well plate at a density of $5 \times 10^5$ cells/mL and cultured overnight to form a monolayer. A scratch wound was created on the monolayer and followed by the laser Irradiation. The cell medium was replaced with the serum-free culture medium, and the scratch wound area was observed and photographed every 24 h under a microscope. The images were analyzed and processed using ImageJ software.

**Fluorescence staining.** After laser Irradiation for 24 h, L929 cells in the experimental and control groups were subjected to fluorescence staining. The cytoskeleton of cells was labeled with rhodamine-labeled phalloidin (Yeasen Biotechnology, Shanghai, China), and the nuclei were labeled with DAPI (Adamas Life, Shanghai, China). The cells were firstly fixed in 4% paraformaldehyde in PBS at room temperature for 10 min, and then were permeabilized with 10% Triton X-PBS for 10 min, and then were stained for the cytoskeleton and nuclei. The images were observed and photographed using an inverted fluorescence microscope (Axio Vert. A1, Carl Zeiss, Jena, Germany) and processed using ZEN and ImageJ software.

**Transcriptome sequencing.** Cells were seeded in a 96-well plate at a density of $1 \times 10^4$ cells/mL. After MIR laser Irradiation, they were cultured for 72 h. The cells were collected and digested with 0.25% trypsin-EDTA (Gibco, NYC, USA). The RNA libraries were sequenced on the Illumina Novaseq™ 6000 platform by OE Biotech, Inc., Shanghai, China. The differential gene expression was analyzed using the criteria of p-value < 0.05 and |log2FC| > 1. Statistics analysis was performed by Wiki Pathways.

**Real Time PCR.** Total RNA was extracted using irzol reagent (GlpBio, CA, USA) and reverse transcribed into cDNA using the SPARKScript II RT Plus Kit (SparkJade, Shandong, China). Real-time PCR was performed with TB Green Premix Ex Taq (TaKaRa) on a LightCycler 480 II real-time PCR system (Roche, Basel, Switzerland) with 40 amplification cycles, and the relative transcript levels were compared by the ΔΔCt method. The primer list is provided in the supplementary table**(Table S1).** GAPDH was used as the internal control. The reactions were conducted in triplicate.

## Associated content

### Supporting Information

The Supporting Information is available free of charge via the Internet at http://pubs.acs.org.

Supplementary Figures and Tables including: Cell viability of fibroblasts after 5.6-μm irradiation for different time (Figure S1); Dependency of cell scratch area on the post-irradiation time after 1-min 5.6-μm irradiation (Figure S2); Photo of MIR irradiation device (Figure S3); Primers used in qPCR (Table S1) (PDF).

## Author information

### Corresponding Author

\* **Feng Zhang** - Quantum Biophotonic Lab, School of Optical-Electrical and Computer-Engineering, University of Shanghai for Science and Technology, Shanghai 200093, China; Wenzhou Institute, University of Chinese Acad-emy of Sciences, Wenzhou 325001, China; orcid.org/0000-0001-6035-4829;

E-mail: fzhang@usst.edu.cn

\* **Liping Wang** - Wenzhou Institute, University of Chinese Academy of Sciences, Wenzhou 325001, China; or-cid.org/0000-0002-9282-2751;

E-mail: lpwang-wiucas@ucas.ac.cn

* **Jun Guo** - Wenzhou Institute, University of Chinese Academy of Sciences, Wenzhou 325001, China.

orcid.org/0000-0001-6944-0731

E-mail: guojun-nbm@wiucas.ac.cn

**Author**

**Xingkun Niu** - Quantum Biophotonic Lab, Key Laboratory of Optical Technology and Instrument for Medicine, Ministry of Education, School of Optical-Electrical and Computer Engineering, University of Shanghai for Science and Technology, Shanghai 200093, China

**Feng Gao** - Wenzhou Institute, University of Chinese Academy of Sciences, Wenzhou 325001, China;

**Shaojie Hou** - The School of Biomedical Engineering, Guangzhou Medical University, Panyu District, Guangzhou 511436, China

**Shihao Liu** - Wenzhou Institute, University of Chinese Academy of Sciences, Wenzhou 325001, China;

**Xinmin Zhao** - Quantum Biophotonic Lab, Key Laboratory of Optical Technology and Instrument for Medicine, Ministry of Education, School of Optical-Electrical and Computer Engineering, University of Shanghai for Science and Technology, Shanghai 200093, China

**Fundings**

This work was supported by the National Key Research and Development Program (No. 2021YFA1200400), the National Natural Science Foundation of China (No. 32271298 and T2241002), and the Wenzhou Institute of the University of Chinese Academy of Sciences (WIUCASQD2021003, WIUCASQD20210011).

**Notes**

The authors declare no conflict of interest.

**Acknowledgements**

We are grateful to OE Biotech. Inc. (Shanghai, China) for assisting in sequencing and bioinformatics analysis.

**Figure legends**

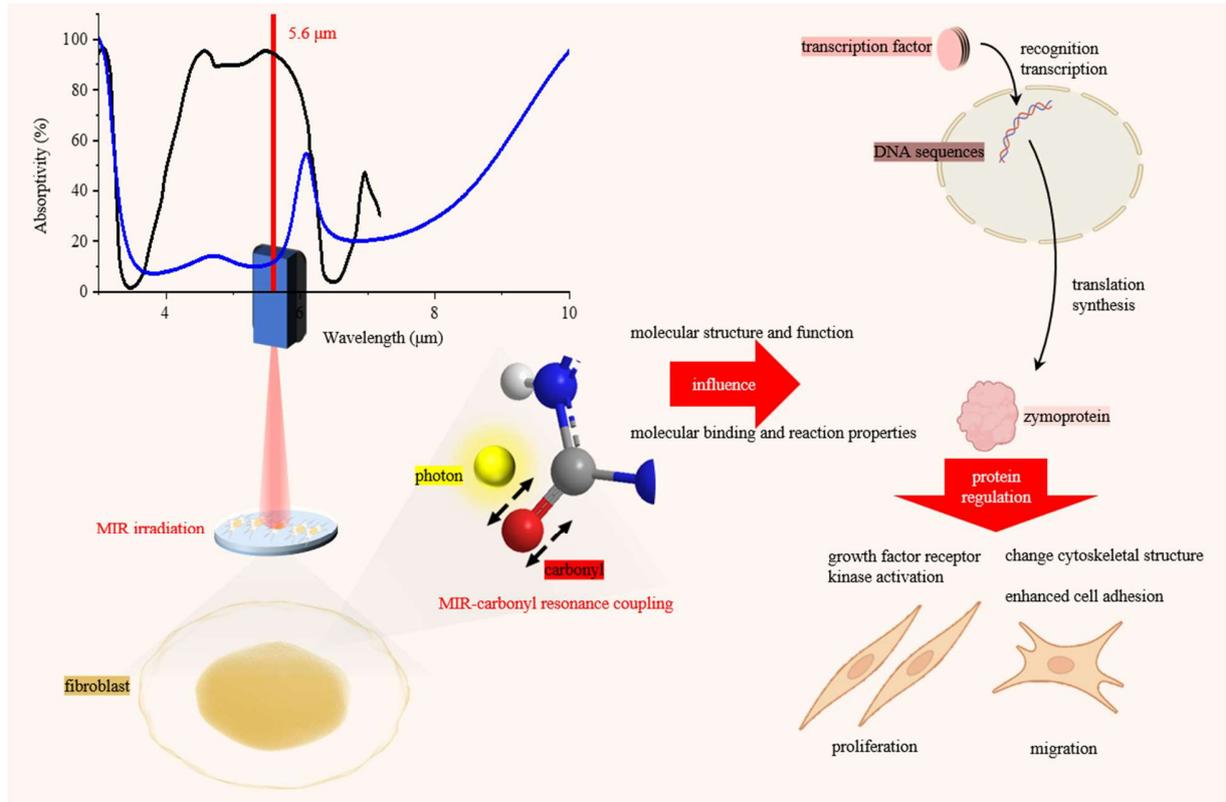

**Figure 1.**

**Schematic illustration of MIR-carbonyl resonance coupling regulating non-thermal effects on fibroblasts** Left panel, 5.6 μm is where the low absorption of water overlaps with the absorption peak of the carbonyl group; and the cell under MIR radiation, the carbonyl group in the cell is vibrationally coupled with photons that match the vibration frequency. Right panel illustrates the process by which carbonyl resonance in cells produces effects that enhance the proliferation and migration capacity of fibroblasts.

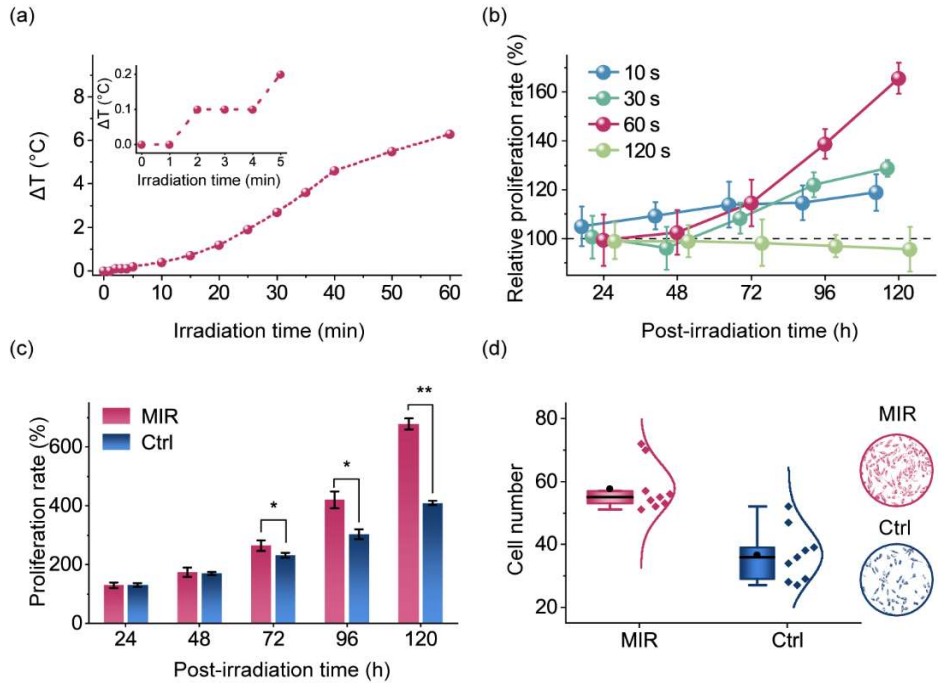

**Figure 2. MIR vibrational coupling enhanced fibroblast proliferation.** (a) Time dose-dependent temperature of cell cultural media with MIR irradiation. Inset is a zoom-in figure. (b) Plots of the relative cell proliferation rate after MIR irradiation with different time against the culture time (hereafter called post-irradiation time). (c) Dependency of cell proliferation rate on the post-radiation time. (d) Comparison of cell numbers after 120-h post-irradiation. Cell photos taken under the microscope are segmented into unit regions of the same size, the average number of cells in unit regions is counted by cell counts for comparison of cell density. On the right is a microscope view of cells at 120h after irradiation, using edge selection and adding color to make the results clearer. The fibroblasts used were L929 cells.

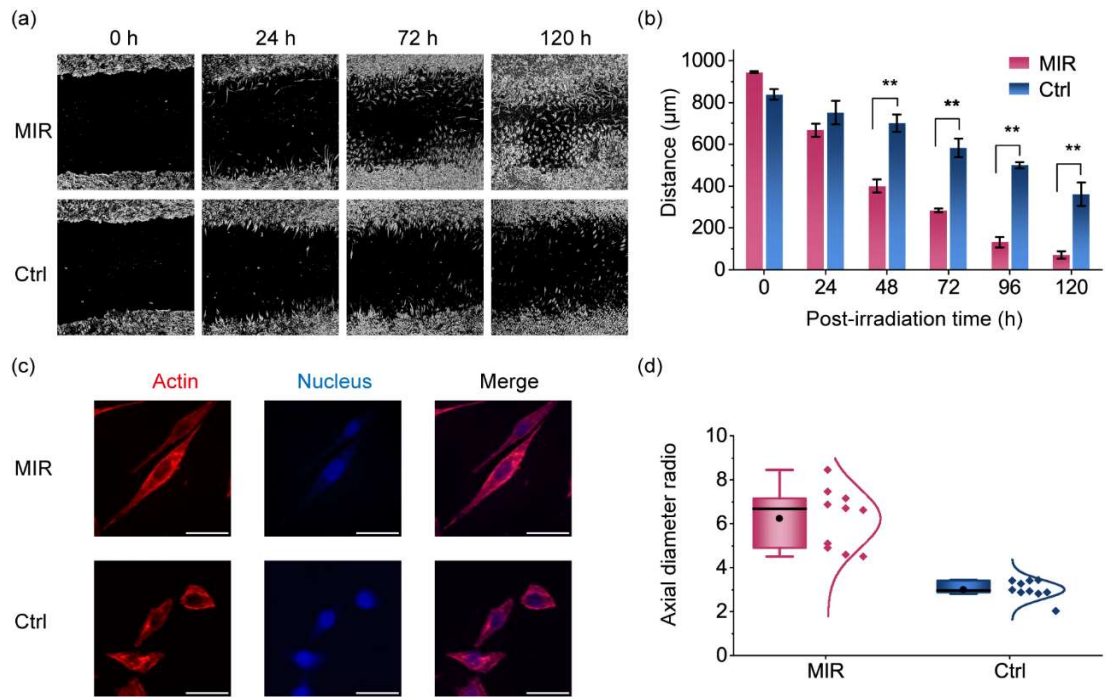

**Figure 3. MIR vibrational coupling enhanced fibroblast migration.** (a) Dependency of cell scratch area on the post-irradiation time after 1-min 5.6-μm irradiation. use edge selections and adding color to make the results clearer. (b) The average width of cell scratch area dependency on the post-Irradiation time. (c) Morphological comparison between fibroblasts at post-Irradiation time of 120 h with and without 1-min 5.6-μm irradiation. (d) Comparison of cellular axial diameter ratio between fibroblasts with and without MIR irradiation. Scale bar represents 10 μm.

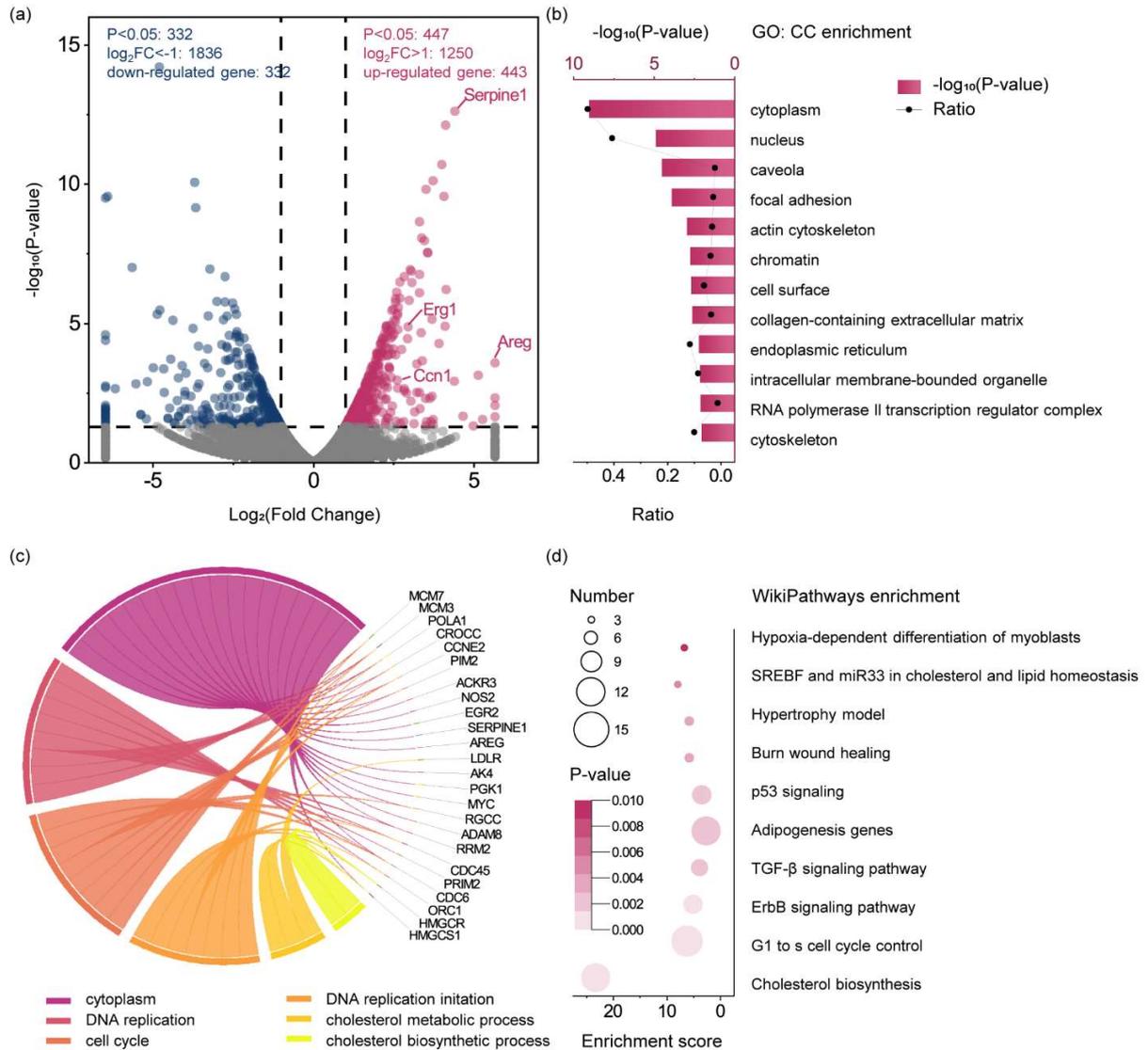

**Figure 4. Transcriptomic sequencing of fibroblasts at 72 h after 1-min 5.6-μm irradiation.** (a) Identification of significantly differentially expressed genes after MIR irradiation. (b) Gene ontology enrichment statistics, the histogram and plot line show the significance and the proportion of differentially expressed genes corresponding to each cell component term, respectively. (c) Chorography displays the interconnectivity between differentially expressed genes and GO enrichment terms. (d) Signaling pathways that could be influenced by MIR irradiation. The color and size of bubbles represent the significance and the number of differentially expressed genes associated with the corresponding pathways, respectively.

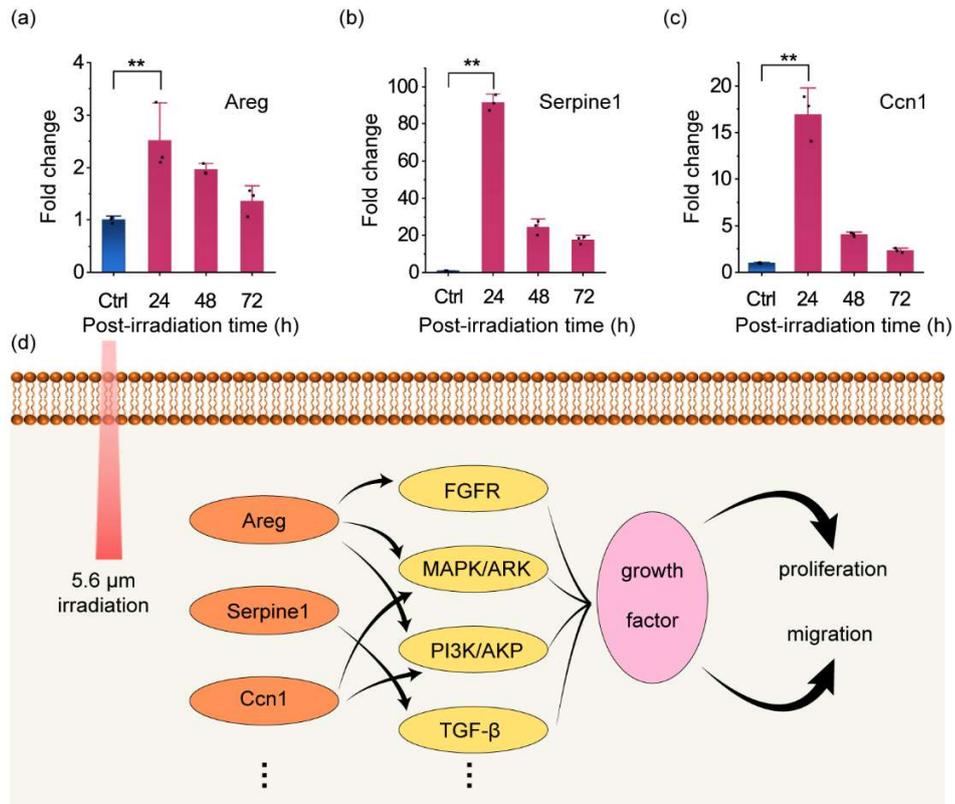

**Figure 5. MIR coupling-induced genetic alterations in fibroblasts at 72 h.** Alterations of (a) Areg, (b) Serpine1 and (c) Ccn1 genes after 1-min 5.6-μm irradiation and extending incubation for another 72 h. (d) Schematic diagram of the correlation among related genes (in orange) and corresponding signal pathways (in yellow) and the growth factors (in pink) in the processes of cell proliferation and migration.

# References

(1) Hertzog, M.; Wang, M.; Mony, J.; Borjesson, K. Strong light-matter interactions: a new direction within chemistry. *Chem Soc Rev* **2019**, *48*, 937-961.

(2) Zhang, X.; Guo, J.; Zhang, F. Mid-Infrared Photons Enhance Polymerase Chain Reaction Efficiency by Strong Coupling with Vibrational DNA Molecules. *ACS photonics* **2023**, *10*, 751-756.

(3) Li, N.; Zhang, F. THz-PCR Based on Resonant Coupling between Middle Infrared and DNA Carbonyl Vibrations. *ACS applied materials & interfaces* **2023**, *15*, 8224-8231.

(4) Liu, X.; Qiao, Z.; Chai, Y.; Zhu, Z.; Wu, K.; Ji, W.; Li, D.; Xiao, Y.; Mao, L.; Chang, C.; Wen, Q.; Song, B.; Shu, Y. Nonthermal and reversible control of neuronal signaling and behavior by midinfrared stimulation. *Proc Natl Acad Sci U S A* **2021**, *118*.

(5) Gotzsche, P. C. Niels Finsen's treatment for lupus vulgaris. *J R Soc Med* **2011**, *104*, 41-2.

(6) Mester A F , M. A. Mester's Method of Laser Biostimulation. *Springer Berlin Heidelberg* **1986**.

(7) Mester, E.; Spiry, T.; Szende, B.; Tota, J. G. Effect of laser rays on wound healing. *Am J Surg* **1971**, *122*, 532-5.

(8) Dompe, C.; Moncrieff, L.; Matys, J.; Grzech-Lesniak, K.; Kocherova, I.; Bryja, A.; Bruska, M.; Dominiak, M.; Mozdziak, P.; Skiba, T. H. I.; Shibli, J. A.; Angelova Volponi, A.; Kempisty, B.; Dyszkiewicz-Konwinska, M. Photobiomodulation-Underlying Mechanism and Clinical Applications. *J Clin Med* **2020**, *9*.

(9) Hamblin, M. R. Photobiomodulation for traumatic brain injury and stroke. *J Neurosci Res* **2018**, *96*, 731-743.

(10) Bensadoun, R. J.; Epstein, J. B.; Nair, R. G.; Barasch, A.; Raber-Durlacher, J. E.; Migliorati, C.; Genot-Klastersky, M. T.; Treister, N.; Arany, P.; Lodewijckx, J.; Robijns, J.; World Association for Laser, T. Safety and efficacy of photobiomodulation therapy in oncology: A systematic review. *Cancer Med* **2020**, *9*, 8279-8300.

(11) Escudero, J. S. B.; Perez, M. G. B.; de Oliveira Rosso, M. P.; Buchaim, D. V.; Pomini, K. T.; Campos, L. M. G.; Audi, M.; Buchaim, R. L. Photobiomodulation therapy (PBMT) in bone repair: A systematic review. *Injury* **2019**, *50*, 1853-1867.

(12) Jansen, E. D.; Echchgadda, I.; Cerna, C. Z.; Sloan, M. A.; Elam, D. P.; Ibey, B. L., Effects of different terahertz frequencies on gene expression in human keratinocytes. In *Optical Interactions with Tissue and Cells XXVI*, 2015.

(13) Lee, D.; Seo, Y.; Kim, Y. W.; Kim, S.; Bae, H.; Choi, J.; Lim, I.; Bang, H.; Kim, J. H.; Ko, J. H. Far-infrared radiation stimulates platelet-derived growth factor mediated skeletal muscle cell migration through extracellular matrix-integrin signaling. *Korean J Physiol Pharmacol* **2019**, *23*, 141-150.

(14) Hsu, Y. H.; Chen, Y. W.; Cheng, C. Y.; Lee, S. L.; Chiu, T. H.; Chen, C. H. Detecting the limits of the biological effects of far-infrared radiation on epithelial cells. *Sci Rep-Uk* **2019**, *9*, 11586.

(15) Chiang, I. N.; Pu, Y. S.; Huang, C. Y.; Young, T. H. Far infrared radiation promotes rabbit renal proximal tubule cell proliferation and functional characteristics, and protects against cisplatin-induced nephrotoxicity. *PLoS One* **2017**, *12*, e0180872.

(16) Zhong, C.; Hou, S.; Zhao, X.; Bai, J.; Wang, Z.; Gao, F.; Guo, J.; Zhang, F. Driving DNA Origami Coassembling by Vibrational Strong Coupling in the Dark. *ACS photonics* **2023**, *10*, 1618-1623.

(17) Gu, K.; Si, Q.; Li, N.; Gao, F.; Wang, L.; Zhang, F. Regulation of Recombinase Polymerase Amplification by Vibrational Strong Coupling of Water. *ACS photonics* **2023**, *10*, 1633-1637.

(18) Gao, F.; Guo, J.; Si, Q.; Wang, L.; Zhang, F.; Yang, F. Modification of ATP hydrolysis by Strong Coupling with O−H Stretching Vibration. *ChemPhotoChem* **2023**, *7*, e202200330.

(19) Bai, J.; Wang, Z.; Zhong, C.; Hou, S.; Lian, J.; Si, Q.; Gao, F.; Zhang, F. Vibrational coupling with O-H stretching increases catalytic efficiency of sucrase in Fabry-Perot microcavity. *Biochem Biophys Res Commun* **2023**, *652*, 31-34.

(20) Sun, L.; Li, Y.; Yu, Y.; Wang, P.; Zhu, S.; Wu, K.; Liu, Y.; Wang, R.; Min, L.; Chang, C. Inhibition of Cancer Cell Migration and Glycolysis by Terahertz Wave Modulation via Altered Chromatin Accessibility. *Research* **2022**, *2022*.

(21) Zhang, J.; Li, S.; Le, W. Advances of terahertz technology in neuroscience: Current status and a future perspective. *Iscience* **2021**, *24*, 103548.

(22) Pollard, T. D.; Cooper, J. A. Actin, a central player in cell shape and movement. *Science* **2009**, *326*, 1208-12.

(23) Burridge K , M. L., Kelly T. Adhesion Plaques: Sites of Transmembrane Interaction between the Extracellular Matrix and the Actin Cytoskeleton. *Journal of Cell Science* **1987**, *8*, 211-229.

(24) Rocnik, E. F.; Chan, B. M.; Pickering, J. G. Evidence for a role of collagen synthesis in arterial smooth muscle cell migration. *J Clin Invest* **1998**, *101*, 1889-98.

(25) Zhang, J.; Guan, M.; Xie, C.; Luo, X.; Zhang, Q.; Xue, Y. Increased growth factors play a role in wound healing promoted by noninvasive oxygen-ozone therapy in diabetic patients with foot ulcers. *Oxid Med Cell Longev* **2014**, *2014*, 273475.

(26) Bakin, A. V.; Tomlinson, A. K.; Bhowmick, N. A.; Moses, H. L.; Arteaga, C. L. Phosphatidylinositol 3-kinase function is required for transforming growth factor beta-mediated epithelial to mesenchymal transition and cell migration. *The Journal of biological chemistry* **2000**, *275*, 36803-10.

(27) Barrientos, S.; Stojadinovic, O.; Golinko, M. S.; Brem, H.; Tomic-Canic, M. Growth factors and cytokines in wound healing. *Wound Repair Regen* **2008**, *16*, 585-601.

(28) Kennedy-Crispin, M.; Billick, E.; Mitsui, H.; Gulati, N.; Fujita, H.; Gilleaudeau, P.; Sullivan-Whalen, M.; Johnson-Huang, L. M.; Suarez-Farinas, M.; Krueger, J. G. Human keratinocytes' response to injury upregulates CCL20 and other genes linking innate and adaptive immunity. *J Invest Dermatol* **2012**, *132*, 105-13.

(29) Cook, P. W.; Mattox, P. A.; Keeble, W. W.; Pittelkow, M. R.; Plowman, G. D.; Shoyab, M.; Adelman, J. P.; Shipley, G. D. A heparin sulfate-regulated human keratinocyte autocrine factor is similar or identical to amphiregulin. *Molecular and Cellular Biology* **1991**, *11*, 2547-2557.

(30) Berasain, C.; Avila, M. A. Amphiregulin. *Seminars in cell & developmental biology* **2014**, *28*, 31-41.

(31) Perbal, B. NOV (nephroblastoma overexpressed) and the CCN family of genes: structural and functional issues. *Mol Pathol* **2001**, *54*, 57-79.

(32) Grzeszkiewicz T M , V. L., Ningyu C , et al. The Angiogenic Factor Cysteine-Rich 61 (CYR61, CCN1) Supports Vascular Smooth Muscle Cell Adhesion and Stimulates Chemotaxis through Integrin α6β1 and Cell Surface Heparan Sulfate Proteoglycans. *Endocrinology* **2002**, *4*, 1441-1450.

(33) Lobel, M.; Bauer, S.; Meisel, C.; Eisenreich, A.; Kudernatsch, R.; Tank, J.; Rauch, U.; Kuhl, U.; Schultheiss, H. P.; Volk, H. D.; Poller, W.; Scheibenbogen, C. CCN1: a novel inflammation-regulated biphasic immune cell migration modulator. *Cell Mol Life Sci* **2012**, *69*, 3101-13.

(34) Moore, R. G.; Brown, A. K.; Miller, M. C.; Skates, S.; Allard, W. J.; Verch, T.; Steinhoff, M.; Messerlian, G.; DiSilvestro, P.; Granai, C. O.; Bast, R. C. The use of multiple novel tumor biomarkers for the detection of ovarian carcinoma in patients with a pelvic mass. *Gynecol Oncol* **2008**, *108*, 402-408.

(35) Simone, T. M.; Higgins, C. E.; Czekay, R. P.; Law, B. K.; Higgins, S. P.; Archambeault, J.; Kutz, S. M.; Higgins, P. J. SERPINE1: A Molecular Switch in the Proliferation-Migration Dichotomy in Wound-"Activated" Keratinocytes. *Adv Wound Care (New Rochelle)* **2014**, *3*, 281-290.

**Table of Contents**

**Enhancing Cell Proliferation and Migration by MIR-Carbonyl Vibrational Coupling: Insights from Transcriptome Profiling**

Xingkun Niu[a], Feng Gao[b], Shaojie Hou[c], Shihao Liu[b], Xinmin Zhao[a], Jun Guo[b,*], Liping Wang[b,*], Feng Zhang[a,b,*]

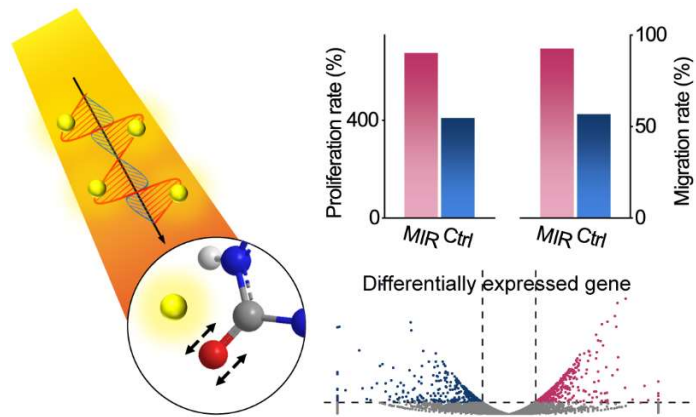

Both proliferation and migration of fibroblast cells can be significantly enhanced by the 5.6-μm laser vibrational strongly coupling with carbonyl groups, which was attributed to the gene selectively differential expression discovered by the transcriptomic profiling.